# The Detection of Interstellar Ethanimine (CH$_3$CHNH) from Observations taken during the GBT PRIMOS Survey

Ryan A. Loomis[1], Daniel P. Zaleski[1], Amanda L. Steber[1], Justin L. Neill[1], Matthew T. Muckle[1], Brent J. Harris[1], Jan M. Hollis[2], Philip R. Jewell[3], Valerio Lattanzi[4], Frank J. Lovas[5], Oscar Martinez, Jr.[4], Michael C. McCarthy[4], Anthony J. Remijan[3], and Brooks H. Pate[1]

[1]*Department of Chemistry, University of Virginia, McCormick Rd, Charlottesville, VA 22904*

[2]*NASA Goddard Space Flight Center, Greenbelt, MD, 20771*

[3]*National Radio Astronomy Observatory, 520 Edgemont Rd., Charlottesville, VA 22904-2475.*

[4]*Harvard-Smithsonian Center for Astrophysics, 60 Garden St., Cambridge, MA 02138, and School of Engineering & Applied Sciences, Harvard University, 29 Oxford St., Cambridge, MA 02138*

[5]*National Institute of Standards and Technology, Gaithersburg, MD 20899*

**Abstract**

We have performed reaction product screening measurements using broadband rotational spectroscopy to identify rotational transition matches between laboratory spectra and the Green Bank Telescope PRIMOS radio astronomy survey spectra in Sagittarius B2 North (Sgr B2(N)). The broadband rotational spectrum of molecules created in an electrical discharge of $CH_3CN$ and $H_2S$ contained several frequency matches to unidentified features in the PRIMOS survey that did not have molecular assignments based on standard radio astronomy spectral catalogs. Several of these transitions are assigned to the *E*- and *Z*-isomers of ethanimine. Global fits of the rotational spectra of these isomers in the range of 8 to 130 GHz have been performed for both isomers using previously published mm-wave spectroscopy measurements and the microwave measurements of the current study. Possible interstellar chemistry formation routes for *E*-ethanimine and *Z*-ethanimine are discussed. The detection of ethanimine is significant because of its possible role in the formation of alanine – one of the twenty amino acids in the genetic code.

# 1. Introduction

The availability of broadband radio astronomy spectral line surveys with high sensitivity and high spectral resolution has greatly aided the identification of new molecular species in astronomical environments and helped advance the understanding of interstellar reaction processes (Belloche et al. 2008, Belloche et al. 2009, Herbst & van Dischoeck 2009, Neill et al. 2012b). Recent technological advances in radio astronomy, in both receiver technology and high-speed digital electronics, have significantly increased observational data throughput. New technologies at facilities such as the Robert C. Byrd Green Bank Telescope (GBT), the Karl G. Jansky Very Large Array (VLA), and the Atacama Large Millimeter/Submillimeter Array (ALMA) allow for high sensitivity to be reached over a wide bandwidth on short time scales (e.g. ~5mK rms over 3.2 GHz bandwidth at Ku band in 15 minutes using the GBT). As a result, the availability of high quality radio astronomy survey spectra is expected to dramatically increase in the next few years. These datasets permit the analysis of column densities of both high and low abundance molecules and the ability to characterize the physical and chemical conditions in astronomical sources on a variety of spatial scales (see e.g. Fortman et al. 2011, Fortman et al. 2012, Neill et al. 2012a). The rapid increase in the volume of observational data presents challenges to the astrochemistry community to develop data and laboratory analysis techniques that are compatible with the new data rates of radio astronomy.

Some of the key technology advances in radio astronomy, especially in the area of high-speed digital signal processing, have also enabled a new generation of laboratory rotational spectroscopy techniques that offer a similar increase in data throughput. Chirped-Pulse Fourier Transform Microwave (CP-FTMW) spectroscopy (Gordon et al. 2008) is a broadband molecular rotational spectroscopy technique that allows deep averaging over large spectral bandwidths (>10 GHz) to yield an experimental "survey spectrum" with high dynamic range (Marshall et al. 2012, Park et al. 2011, Steber et al. 2011). This technology allows laboratory spectroscopy to move beyond the traditional "targeted search" model where the focus is on the production and spectroscopic characterization of a single chemical species of possible astrochemical interest. Instead, it is feasible to perform "reaction product screening" experiments where high abundance interstellar molecules are subjected to reactive conditions and the complex mixture of product molecules is simultaneously characterized by rotational spectroscopy. The combination of broadband spectral surveys from radio astronomy and the laboratory provide a new way to mine these data sets and identify new interstellar molecules through rapid identification of spectral overlaps in the laboratory with the numerous unassigned spectral features in the radio astronomy spectra (Remijan et al. 2009).

The GBT PRebiotic Interstellar MOlecule Survey (PRIMOS) Legacy Project was undertaken in 2008 to provide complete spectral line data between 300 MHz and 50 GHz toward the Sgr B2(N) pointing position – the preeminent region in the Galaxy for molecular line studies (see e.g. Nummelin et al. 1998, Turner 1991, Cummins et al. 1986 and references therein)[1]. The broadband rotational spectrum of the mixture of molecules produced from an electrical discharge of $CH_3CN$ and $H_2S$ (Zaleski et al 2012b, McCarthy et al. 2000) contained several matches to unidentified features in the GBT PRIMOS data of Sgr B2(N) that did not have molecular assignments in the radio astronomy spectral catalogs. These frequency coincident overlapping u-line features were subsequently identified as *E*- and *Z*-ethanimine ($CH_3CHNH$) (this work)[2]. Ethanimine was suggested as a possible interstellar molecule candidate after the interstellar detection of methanimine ($CH_2NH$) in 1973 (Godfrey 1973) and microwave and mm-wave spectra of its two conformers, *E*- and *Z*-ethanimine, were characterized in order to guide astronomical searches (Lovas et al. 1980; Brown et al. 1980). The formation and reaction chemistry of aldimines such as ethanimine is also of special interest to the formation of amino acids and other prebiotic molecules (Arnaud et al. 2000, Basiuk et al. 2001, Danger et al. 2011, Koch et al. 2008, Woon 2002).

---

[1] Access to the entire PRIMOS dataset and specifics on the observing strategy including the overall frequency coverage, is available at http://www.cv.nrao.edu/~aremijan/PRIMOS/

[2] *E*- and *Z*- refer to the relative positions of the -NH hydrogen and the methyl group about the double bond. See sketches of structures in Lovas et al. 1980.

## 2. Observations and Laboratory Measurements

Transitions of *E*- and *Z*-ethanimine were identified in data observed as part of the PRIMOS survey, a NRAO key project from 2008 January through 2011 July. An LSR source velocity of +64 km s$^{-1}$ was assumed and antenna temperatures were recorded on the $T_a^*$ scale (Ulich & Haas 1976) with estimated 20% uncertainties. Data were taken in position-switching mode, with two minute scans toward the on position ($\alpha_{J2000} = 17^h47^m19^s.8$, $\delta_{J2000} = -28°22'17.0"$) and towards the off position, 1° east in azimuth. For more details on the PRIMOS survey observations see e.g. Neill et al. (2012a) . The data reduction and line profile fitting was completed using the GBTIDL reduction software.

New laboratory measurements of the rotation-torsion spectrum of *E*- and *Z*-ethanimine were taken at microwave frequencies to extend the frequency coverage from previously published millimeter wave data. Initial broadband measurements of a $CH_3CN$ and $H_2S$ electrical discharge (McCarthy et al. 2000) were performed using CP-FTMW spectroscopy in the frequency ranges of 6.5-18.5 GHz (Gordon et al. 2008) and 25-40 GHz (Zaleski et al. 2012a). Features in this dataset were assigned using spectroscopic catalogues, with unidentified features then compared against ab-initio predictions of the rotational spectra of numerous molecules. The microwave transitions of ethanimine were measured at higher spectral resolution at the Harvard

Smithsonian Center for Astrophysics between 8-40 GHz with uncertainties of 5 kHz on a Flygare FTMW spectrometer (Grabow et al. 2005) to fully resolve the nuclear quadrupole hyperfine and methyl group internal rotation fine structure. Additional measurements from 50-90 GHz were made using microwave-microwave double resonance (Nakajima, Sumiyoshi, & Endo 2002) with uncertainties of 20 kHz. A global fit to previously published millimeter wave (Lovas et al. 1980) and new microwave laboratory data was performed for each species using the extended Inertial Axis Method implemented in the XIAM program[3] (Hartwig & Dreizler 1996). Predicted frequencies from this fit have been added to the SLAIM database. Rest frequencies from the global fits are listed in Tables 1a & b, with standard deviations of 12 kHz and 56 kHz for the *E*- and *Z*-ethanimine fits, respectively.

## 3. Results and Probability of Detection

Twenty low energy, high line strength transitions of *E*- and *Z*-ethanimine were identified in the PRIMOS data between 13 and 47 GHz (Figures 1-3): 12 transitions of *E*-ethanimine were detected, both a-type and b-type ($\mu_a$ = 2.78(8) x$10^{-30}$ Cm, [ $\mu_a$ = 0.834(23) D], $\mu_b$ = 6.278(17) x$10^{-30}$ Cm [$\mu_b$ = 1.882(5) D]) (Lovas et al. 1980), while only 8 a-type transitions of *Z*-ethanimine were detected ($\mu_a$ = 7.939(7) x$10^{-30}$ [$\mu_a$ = 2.380(2) D], $\mu_b$ = 1.45(31) x$10^{-30}$ Cm [$\mu_b$ = 0.445(93) D]) (Brown 1980), due to the low quantum mechanical line strength of its b-type transitions. This encompasses all strong transitions within PRIMOS frequency coverage predicted at a characteristic excitation temperature of 10 K to be above the measured noise limit in the PRIMOS spectrum. Of these detected transitions, all exhibit the expected A/E splitting and hyperfine splitting was clearly resolved in 4 transitions. No transitions are missing, although both the A and E $2_{12}$-$1_{11}$ transitions of *Z*-ethanimine are partially blended with the $3_{03}$-$2_{02}$ A transition of $CH_3OCHO$ $v_t=1$ and the $4_{23}$-$3_{22}$ transition of $CH_3CH_2CN$ v=0, respectively (Figures 3c,d). The A and E $2_{11}$-$2_{02}$ transitions of *E*-ethanimine are also partially blended with the 5-4 transition of $HC_3N$ $v_6=1$ and $HC_3N$ v=0, respectively (Figures 2c,d).

---
[3] XIAM is available online at http://www.ifpan.edu.pl/~kisiel/introt/introt.htm#xiam

Observations of the *E*- and *Z*-ethanimine transitions are presented in Figures 1, 2, and 3. All detected transitions were observed at a nominal rest velocity of $v_{LSR}$ = +64 km s$^{-1}$, with transitions in Figures 1(a-d), 2(a,b,e,f), and 3b,f showing an additional $v_{LSR}$ = +82 km s$^{-1}$ component, consistent with other molecular transitions detected with the GBT (see, e.g. Hollis et al. 2004, Remijan et al, 2008 and references therein). In Figures 3c and 3d, the nearby emission features preclude detection of the +82 km s$^{-1}$ component and partially blend with the +64 km s$^{-1}$ velocity component, reducing the fitted intensity and line widths. All transitions, except for the $3_{03}$–$2_{12}$ and $4_{04}$-$3_{13}$ transitions of *E*-ethanimine, were observed in absorption (discussed further in Section 4.1).

All detected transitions of *E*- and *Z*-ethanimine, including best-fit Gaussian line widths and peak intensities, are reported in Tables 1a and 1b, respectively. When resolved, the widths of all hyperfine components were fixed by the Gaussian width of the strongest line. For unresolved hyperfine components, a single Gaussian profile was fit to the whole feature thereby artificially broadening the measured line width (see Table 1). In addition, the superposition of several of the hyperfine components from the +64 and +82 km s$^{-1}$ velocity components in Figures 1a,b and 3a resulted in strong coupling between their intensities when fitting, explaining the large uncertainties reported in Table 1. Features with resolvable hyperfine structure had line widths of ~10-15 km s$^{-1}$, comparable to other nitriles and nitrile derivatives observed toward Sgr B2(N) (Remijan et al. 2005)

All features identified as ethanimine transitions fall within 0.05 MHz of predicted rest frequencies, matching the predicted A/E splittings and hyperfine splittings (where resolved). In addition, hyperfine features have relative intensities that match those predicted by quantum mechanical calculations, and all transitions identified as A/E pairs match each other in intensity within calibration uncertainties (<20%). Based on the line density of the PRIMOS survey (approximately 4 features/100 MHz), it is unlikely that all features identified as transitions of ethanimine could be explained by random unidentified features coincident with ethanimine rest frequencies (see argument outlined in Neill et al. 2012a). The transitions from the GBT PRIMOS data unambiguously establish the presence of both the *E*- and *Z*- isomers of ethanimine towards SgrB2(N).

## 4. Discussion

4.1 *Temperature and Column Density Determination*

The excitation temperature and total beam-averaged column density for *E*- and *Z*-ethanimine were determined using the formalism described in Remijan et al. (2005) for all measured absorption line features, assuming optically thin emission and a beam filling factor of unity. Using the spectral line parameters given in Table 1, a best fit temperature of 6(2) K and column density of 2.3(5) x $10^{13}$ cm$^{-2}$ were determined for *Z*-ethanimine by a linear least squares method. This cold rotational temperature is typical of other nitriles and nitrile derivatives in Sgr B2(N) (Zaleski et al. 2012b, Remijan et al. 2005) and possibly indicates sub-thermal behavior, where the rotational temperature observed is lower than the kinetic temperature.

Although *Z*-ethanimine was best fit with this model, no single temperature was able to fit all features of *E*-ethanimine. This behavior may be partially explained by a failure of the assumptions of beam filling and a local continuum temperature equal to that measured by the GBT[4]. However, these explanations do not account for the presence of some *E*-ethanimine transitions in emission while others are in absorption. The $3_{03}$-$2_{12}$ and $4_{04}$-$3_{13}$ transitions of *E*-ethanimine observed in emission are the only b-type transitions with $K_a = 0$ in the excited state, while all other b-type transitions have $K_a = 1$ in the excited state, suggesting a higher excitation temperatures for the $\Delta K_a = +1$ transitions, or a possible population inversion. If a pumping mechanism (either collisional or radiative) is responsible for a population inversion, or the excitation temperatures differ, a local thermal equilibrium (LTE) model such as the one used would not fit the data.

Using a fixed temperature of 6 K for *E*-ethanimine, a column density of ~7 x $10^{13}$ cm$^{-2}$ was determined for the transitions in absorption, which is similar to the Z-ethanimine column density. The observation of both isomers is consistent with formation chemistry under kinetic control. *Ab initio* calculations at MP2/6-311++G(d,p) level of theory (using Gaussian09W, Frisch et al. 2009) give *E*-ethanimine as the lower energy conformer by 4.24 kJ/mol (510 K) with a barrier to isomerization to the less stable *Z*-ethanimine of 134 kJ/mol. Once formed, it would be unlikely that *Z*-ethanimine could isomerize to the more stable form under interstellar conditions.

4.2 *Formation Mechanism*

There is strong experimental evidence that ethanimine can be produced on ices from its parent nitrile, $CH_3CN$. The reaction mechanism, which converts a parent nitrile (R-CN) to its corresponding primary aldimine (R-CH=NH), involves sequential hydrogen atom addition to the nitrile, and has previously been proposed as a possible formation route for methanimine ($CH_2NH$) (Dickens et al. 1997, Woon 2002). The first hydrogen atom addition in this reaction has a barrier, which has been calculated for the formation of methanimine to be 7.3 kJ/mol (873 K) (Woon 2002). The second hydrogen atom addition is a radical-radical reaction which is expected to proceed with essentially no barrier.

---

[4] Measured continuum temperatures for the Sgr B2(N) region as measured by the GBT from the PRIMOS data can be found in Hollis et al. (2007).

A number of studies have provided experimental evidence for such a mechanism. A previous study of hydrogen atom additions to $CH_3CN$ in ices at 77 K identified the production of the radical intermediate, $CH_3CH=N\bullet$ (Svejda and Volman 1970). In this experiment, hydrogen atoms were created within the ice by photodissociating co-deposited HI with UV excitation, and electron paramagnetic resonance spectroscopy was employed, which exclusively detects the presence of the radical. Other recent studies with a focus on interstellar ice chemistry have shown that sequential hydrogen atom additions occur for both HCN and $CH_3CN$ (Theule et al. 2011, Hudson et al. 2008), with only the final fully hydrogenated products observed ($R-CH_2-NH_2$, with R = H for HCN and R = $CH_3$ for $CH_3CN$ as the parent nitriles). These results show that the hydrogen atom addition reactions that convert nitriles to primary aldimines are feasible, but also that the final distribution of products will be significantly affected by the amount of atomic hydrogen present.

4.3 *Implications for Prebiotic Chemistry*

The identification of synthetic methods to produce amino acids from simple chemical precursors is an active area of astrobiology. Experiments using UV radiation of simple ices composed of high abundance interstellar molecules have shown that amino acids can be formed and that there are likely several responsible mechanisms (Bernstein et al. 2002, Munoz Caro et al. 2002). Many studies have tested the feasibility of a synthetic route in interstellar ices based on the well-known Strecker synthesis of amino acids from aldehydes and ketones in the presence of ammonia, HCN, and water (Elsila et al. 2007, and references therein). Amino acids have been detected following hydrolysis of UV-irradiated model interstellar ices containing the precursors necessary for a Strecker-like synthesis, with methanol converted to an aldehyde as the first step of the route, as shown in the top panel of Figure 4 (Bernstein et al. 1995).

It has been shown, however, that the presence of alcohols in ices is not necessary for amino acid production (Elsila et al. 2007). Additional measurements in this study using isotopic labeling showed that the major final product is inconsistent with the Strecker synthesis, with the amino nitrogen of the major product not derived from $NH_3$ as required in the Strecker synthesis. Woon (2002) presented a proposal for amino acid synthesis using sequential hydrogen atom addition to the nitrile to form an aldimine, with a further second hydrogen atom addition step to form an amine radical. A final radical-radical reaction with COOH (from formic acid – a known interstellar species) would create the amino acid, as shown in the middle panel of Figure 4. Elsila *et al.* (2007), however, point out that the second step of the Woon proposal is also inconsistent with their isotopic data, suggesting instead that if the primary aldimine produced in the Woon mechanism undergoes HCN addition to create the amino-nitrile, then the resultant amino acid has an isotopic labeling pattern consistent with experiment, shown in the bottom panel of Figure 4.

In this mechanism for amino acid formation, the key step is the formation of a primary aldimine that can be co-deposited in the interstellar ice with HCN to permit subsequent formation of the aminonitrile precursor. As the amino acids in the experimental work are not observed directly in the ices but rather only after the subsequent hydrolysis step, the aminonitriles may serve as a reservoir species. This mechanism for amino acid formation has the advantage that the aminonitrile is more photostable than the corresponding amino acid, increasing the chances that it would be delivered to a newly formed planet for subsequent conversion to the amino acid by hydrolysis (Chyba and Sagan 1992, Bernstein et al. 2004). For the formation of the simplest amino acid, glycine, interstellar production of the two key species (methanimine and aminoacetonitrile) has been confirmed by radio astronomy detection (Godfrey 1973, Belloche et al. 2008). The present detection of ethanimine has identified the first reaction intermediate required for the formation of alanine.

## 5. Conclusions

The spectroscopic evidence supporting the interstellar detection of *E*- and *Z*-ethanimine in the GBT PRIMOS observations of Sgr B2(N) from laboratory reaction product screening measurements has been presented. A column density on the order of $10^{13}$ cm$^{-2}$ has been determined for both species, and a rotational temperature of 6 K has been fit for *Z*-ethanimine. *E*-ethanimine shows clear non-LTE behavior, and collisional cross-sections are necessary to more accurately determine the abundance and excitation of this species via radiative transfer models in future studies. The interstellar formation chemistry of ethanimine may involve sequential hydrogen atom addition to the parent nitrile (CH$_3$CN) in interstellar ices – a mechanism that has strong support in studies of ice chemistry. The formation of R-CH=NH species in interstellar ices containing nitriles has also been proposed as an intermediate step in forming stable precursors to amino acids that can be formed by subsequent hydrolysis reactions in a warm environment. The detection of ethanimine establishes an interstellar precursor to the formation of alanine – the second most common amino acid in the primary sequence of proteins (Doolittle 1989).

**6. Acknowledgements**

The authors acknowledge support from the Centers for Chemical Innovation program of the National Science Foundation (CHE-0847919), NSF Chemistry (CHE-1213200), and the College Science Scholars program at the University of Virginia. We also thank J. Corby, B. McGuire, and R. Pulliam for helpful comments to improve the manuscript. Finally, we thank an anonymous referee for valuable comments and suggestions on the manuscript. The National Radio Astronomy Observatory is a facility of the National Science Foundation operated under cooperative agreement by Associated Universities, Inc.

Table 1. Observed transitions for the A- and E-torsional sublevels of *E*- and Z-ethanimine in the ground vibrational state.

| Conformer | Transition A/E $J'_{kk} - J''_{kk}$ | F' - F'' | Frequency[a,b] (MHz) | $E_u$ (K) | $S_{ij}\mu^2$ (D$^2$) | + 64 km s$^{-1}$ $\Delta T_a^{*c}$ (mK) | $\Delta V^{cd}$ (km s$^{-1}$) | + 82 km s$^{-1}$ $\Delta T_a^{*c}$ (mK) | $\Delta V^{cd}$ (km s$^{-1}$) |
|---|---|---|---|---|---|---|---|---|---|
| E | (A) $3_{03} - 2_{12}$ | 2 - 1 | 13026.113 | 5.32 | 2.28 | 7.8(-)[e] | 15.3(0.5) | 7.0(1.1) | 15.3(0.5) |
|   |   | 4 - 3 | 13026.749 | 5.32 | 4.89 | 38.4(1.4) | 15.3(0.5) | 33.3(-)[e] | 15.3(0.5) |
|   |   | 3 - 2 | 13027.992 | 5.32 | 3.38 | 29.8(0.9) | 15.3(0.5) | 24.3(2.0) | 15.3(0.5) |
| E | (E) $3_{03} - 2_{12}$ | 2 - 1 | 13104.146 | 5.32 | 2.26 | 17.9(-)[e] | 11.9(0.4) | 16.9(0.9) | 11.9(0.4) |
|   |   | 4 - 3 | 13104.780 | 5.32 | 4.83 | 33.1(1.9) | 11.9(0.4) | 35.1(-)[e] | 11.9(0.4) |
|   |   | 3 - 2 | 13106.019 | 5.32 | 3.34 | 24.4(0.8) | 11.9(0.4) | 24.6(1.9) | 11.9(0.4) |
| E | (A) $1_{01} - 0_{00}$ | 2 - 1 | 18480.081 | 0.89 | 1.16 | -9.6(1.6) | 10.0(0.2) | blended | |
| E | (E) $1_{01} - 0_{00}$ | 2 - 1 | 18479.545 | 0.89 | 1.16 | -11.4(3.7) | 10.0(0.2) | blended | |
| E | (A) $1_{10} - 1_{01}$ | 0 - 1 | 44467.542 | 3.02 | 1.77 |  |  |  |  |
|   |   | 2 - 2 | 44469.216 | 3.02 | 6.63 | -19.7(1.6) | 18.4(1.3) | -14.0(2.1) | 6.7(0.8) |
|   |   | 1 - 2 | 44470.130 | 3.02 | 2.21 |  |  |  |  |
| E | (E) $1_{10} - 1_{01}$ | 0 - 1 | 44407.851 | 3.02 | 1.77 |  |  |  |  |
|   |   | 2 - 2 | 44409.478 | 3.02 | 6.65 | -18.9(1.0) | 15.6(0.9) | -11.9(1.4) | 11.8(1.4) |
|   |   | 1 - 2 | 44410.361 | 3.02 | 2.21 |  |  |  |  |
| E | (A) $2_{11} - 2_{02}$ | 1 - 1 | 45566.104 | 4.85 | 3.93 |  |  |  |  |
|   |   | 3 - 3 | 45566.664 | 4.85 | 10.87 | -19.9(2.8) | 11.0(1.2) | blended | |
|   |   | 2 - 2 | 45567.671 | 4.85 | 6.07 |  |  |  |  |
| E | (E) $2_{11} - 2_{02}$ | 1 - 1 | 45495.409 | 4.84 | 3.94 |  |  |  |  |
|   |   | 3 - 3 | 45495.967 | 4.84 | 10.90 | -32.8(1.5) | 13.2(0.7) | blended | |
|   |   | 2 - 2 | 45496.971 | 4.84 | 6.08 |  |  |  |  |
| E | (A) $3_{12} - 3_{03}$ | 4 - 4 | 47249.906 | 7.59 | 14.47 | -37.1(1.6) | 16.4(0.8) | -31.1(2.8) | 10.0(0.7) |
|   |   | 3 - 3 | 47251.069 | 7.59 | 10.09 |  |  |  |  |
| E | (E) $3_{12} - 3_{03}$ | 4 - 4 | 47176.006 | 7.58 | 14.51 | -52.3(1.7) | 20.2(0.7) | -28.3(2.3) | 9.7(0.6) |
|   |   | 3 - 3 | 47177.169 | 7.58 | 10.11 |  |  |  |  |
| Z | (A) $1_{01} - 0_{00}$ | 1 - 1 | 18478.046 | 0.89 | 5.98 | -13.6(0.9) | 10.1(0.2) | blended | |
|   |   | 2 - 1 | 18479.160 | 0.89 | 9.97 | -35.5(1.3) | 10.1(0.2) |  |  |
| Z | (E) $1_{01} - 0_{00}$ | 1 - 1 | 18477.268 | 0.89 | 5.98 | -17.8(0.8) | 10.1(0.2) | blended | |

| | | 2 - 1 | 18478.413 | 0.89 | 9.97 | -31.8(1) | 10.1(0.2) | | |
|---|---|---|---|---|---|---|---|---|---|
| Z | (A) $2_{12} - 1_{11}$ | 2 - 1 | 35778.075 | 4.53 | 6.74 | | | | |
| | | 3 - 2 | 35779.255 | 4.53 | 12.58 | -32.1(0.8) | 13.3(0.4) | blended | |
| | | 1 - 0 | 35780.083 | 4.53 | 2.99 | | | | |
| Z | (E) $2_{12} - 1_{11}$ | 2 - 1 | 35794.647 | 4.53 | 6.73 | | | | |
| | | 3 - 2 | 35795.865 | 4.53 | 12.57 | -30.2(1.9) | 17.5(0.7) | blended | |
| | | 1 - 0 | 35796.700 | 4.53 | 2.99 | | | | |
| Z | (A) $2_{02} - 1_{01}$ | 1 - 0 | 36931.311 | 2.66 | 3.99 | -11.4(-)$^e$ | 13.3(0.6) | | |
| | | 2 - 1 | 36932.222 | 2.66 | 8.97 | -118(8) | 13.3(0.6) | -36.2(-)$^e$ | 13.3(0.6) |
| | | 3 - 2 | 36932.313 | 2.66 | 16.75 | | | | |
| Z | (E) $2_{02} - 1_{01}$ | 1 - 0 | 36929.961 | 2.66 | 3.99 | -23.5(-)$^e$ | 13.3(0.6) | | |
| | | 2 - 1 | 36930.703 | 2.66 | 8.97 | -98(20) | 13.3(0.6) | -31.5(6) | 13.3(0.6) |
| | | 3 - 2 | 36930.793 | 2.66 | 16.75 | | | | |
| Z | (A) $2_{11} - 1_{10}$ | 2 - 1 | 38135.387 | 4.70 | 6.74 | | | | |
| | | 3 - 2 | 38136.514 | 4.70 | 12.59 | -17.5(0.8) | 12.8(0.7) | <20 (2σ) | ••• |
| | | 1 - 0 | 38138.021 | 4.70 | 3.00 | | | | |
| Z | (E) $2_{11} - 1_{10}$ | 2 - 1 | 38116.348 | 4.70 | 6.75 | | | | |
| | | 3 - 2 | 38117.461 | 4.70 | 12.60 | -36.5(2.5) | 26.0(0.6) | <20 (2σ) | ••• |
| | | 1 - 0 | 38118.968 | 4.70 | 3.00 | | | | |

$^a$ Beam sizes, efficiencies, and continuum temperatures for each respective frequency can be found in Hollis et al. 2007

$^b$ Rest frequencies are calculated from global fits of all laboratory measurements with Type A, $k = 2$ (2 sigma) standard deviations of 24 kHz for *E*-ethanimine and 112 kHz for *Z*-ethanimine. All lines have been experimentally measured with (*Obs – Calc*) frequency uncertainties of ≤ 25 kHz (Taylor & Kuyatt 1994).

$^c$ The uncertainties for the intensities and line widths are Type B, $k = 1$ (1 sigma) (Taylor & Kuyatt 1994)

$^d$ Line widths were fixed for individual hyperfine components of a transition during fitting.

$^e$ Due to blended hyperfine components, these line intensities are coupled and have high uncertainties.

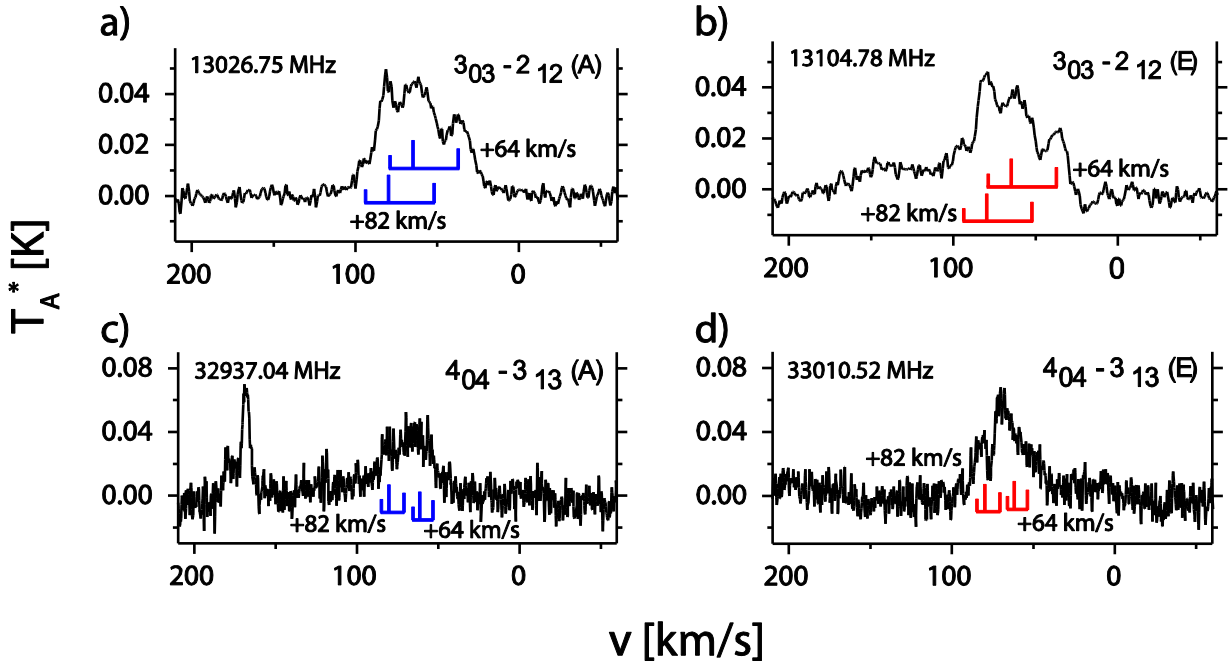

Fig. 1. Interstellar spectra from the GBT PRIMOS survey for *E*-ethanimine transitions. Relative intensity markers are shown where hyperfine splitting is expected, with the A torsional state in blue and the E torsional state in red.

# E - CH$_3$CHNH

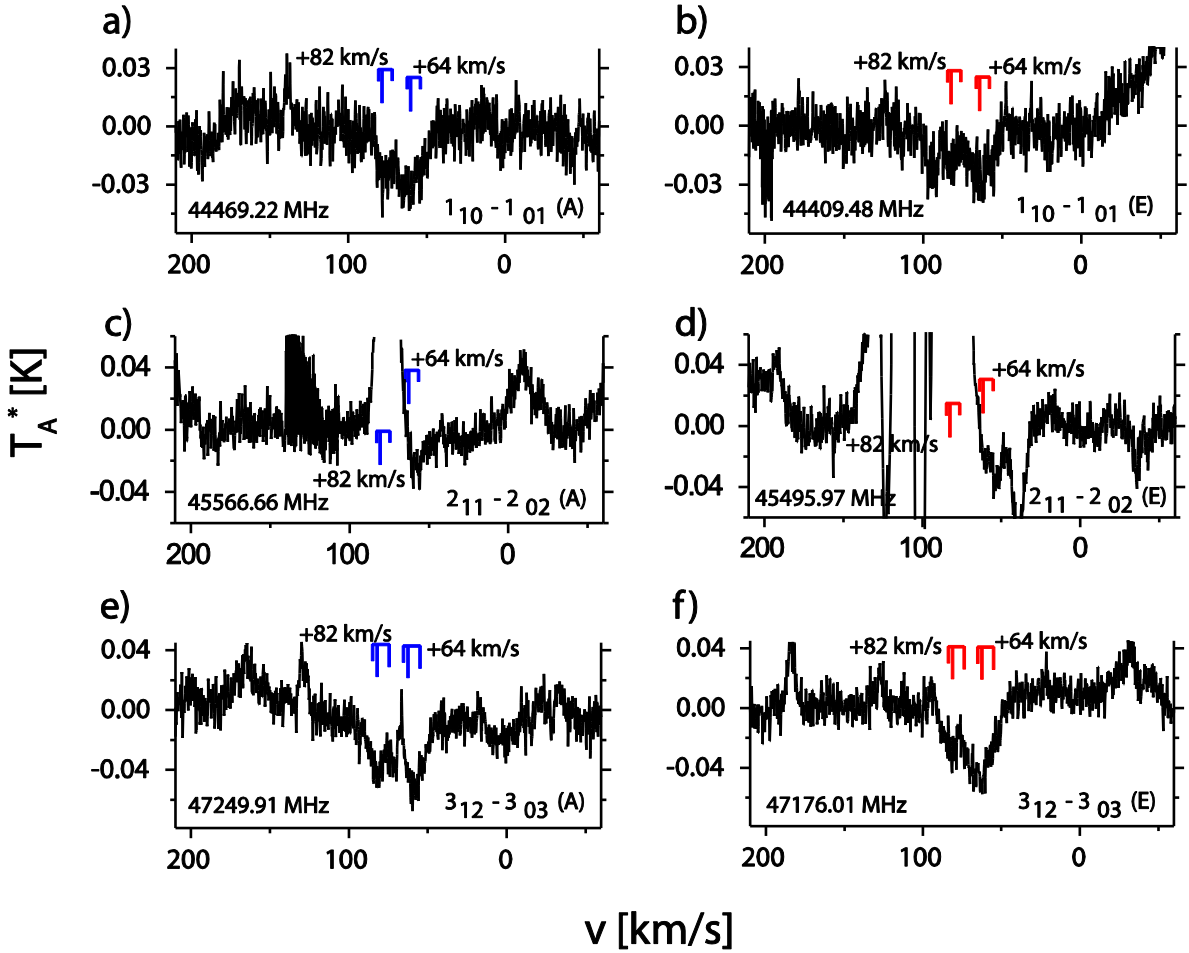

Fig. 2. Interstellar spectra from the GBT PRIMOS survey for *E*-ethanimine transitions. Relative intensity markers are shown where hyperfine splitting is expected, with the A torsional state in blue and the E torsional state in red. In panel e, the +64 km s$^{-1}$ velocity component in absorption is blended with $4_{04}$-$3_{03}$ E transition of CH$_3$OCHO v=1 in emission at 47249.49 MHz.

# Z - CH$_3$CHNH

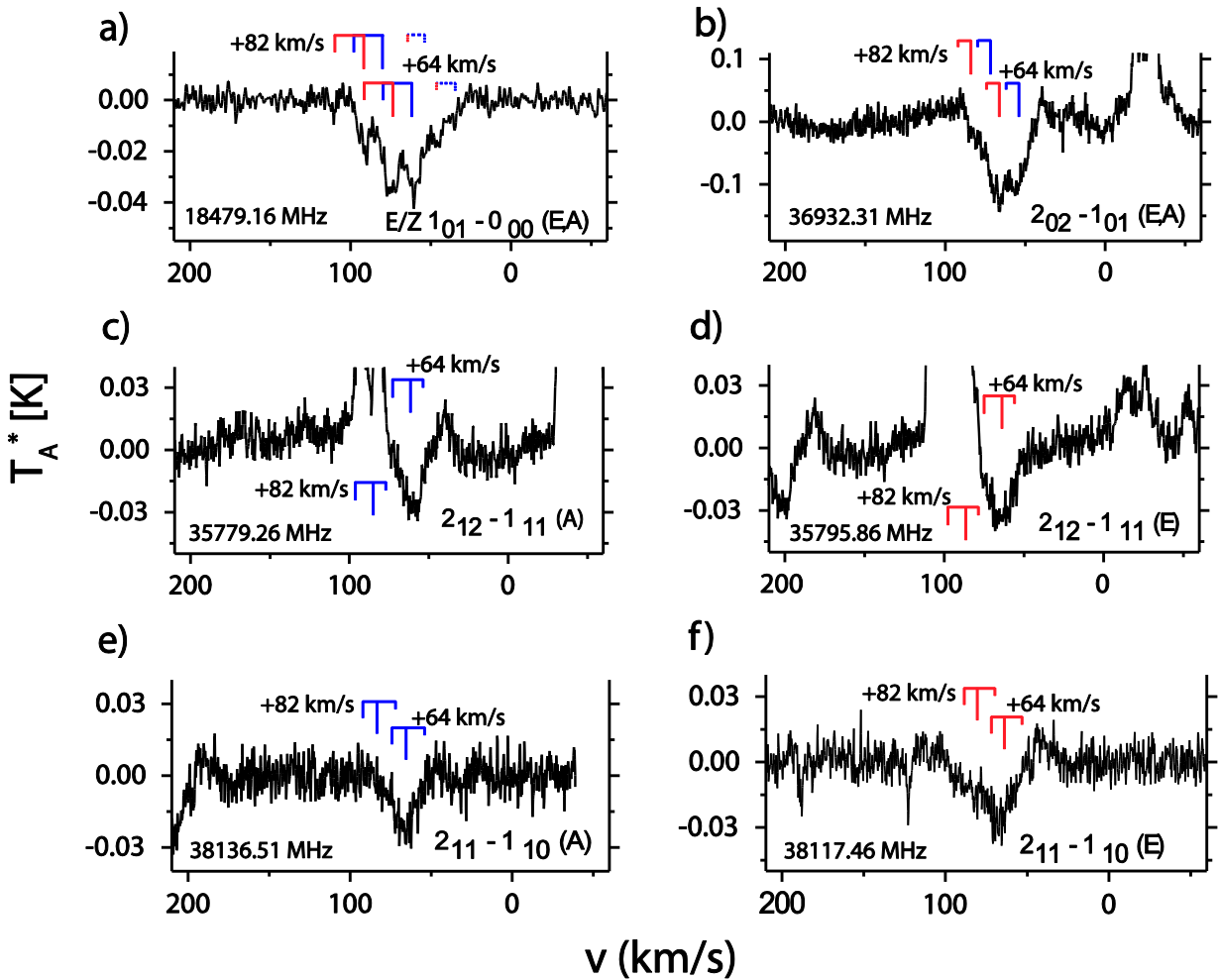

Fig. 3. Interstellar spectra from the GBT PRIMOS survey for Z-ethanimine transitions. Relative intensity markers are shown where hyperfine splitting is expected, with the A torsional state in blue and the E torsional state in red. Panel a additionally shows a weak contribution from the $1_{01} - 0_{00}$ a-type transition E-ethanimine that is on the high frequency side of the transition. The calculated spectrum for this overlapping E-ethanimine transition is shown with dashed lines.

## Bernstein Strecker:

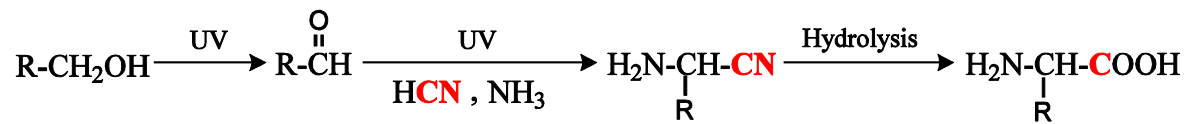

## Woon Radical-Radical:

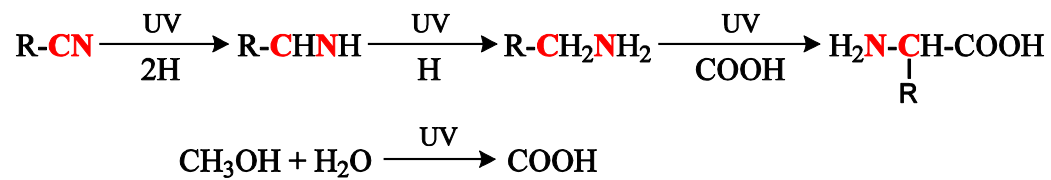

## Elsila Modified Nitrile:

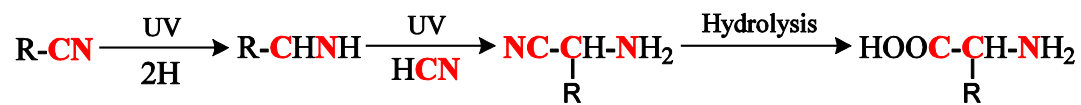

Fig 4. Possible formation routes of amino acids, showing isotopically labeled atoms from Elsila et al. (2007) in bold red.